\documentclass[aps,amssymb,floatfix,prd,amsmath,preprintnumbers]{revtex4}
\RequirePackage{float}
\usepackage{epstopdf}
\usepackage{hyperref}
\usepackage{capt-of}
\usepackage{graphicx}  
\usepackage{dcolumn}   
\usepackage{bm}
\usepackage{booktabs}
\usepackage{nicefrac} 
\usepackage{mathtools}

\begin{document}
	
\title{Interacting Tsallis and R\'enyi Holographic Dark Energy with Hybrid Expansion Law}

\author{Snehasish Bhattacharjee \footnote{Email: snehasish.bhattacharjee.666@gmail.com \\ ORCID: https://orcid.org/0000-0002-7350-7043 \\}
  }\
  
\affiliation{Department of Astronomy, Osmania University, Hyderabad-500007,
India}

\begin{abstract}
The manuscript presents the dynamics of Tsallis holographic dark energy (THDE) and R\'enyi holographic dark energy (RHDE) prescribed by a non-linear interaction in the FRW spacetime and for a scale factor evolving with a composite power law-exponential (hybrid) form. To construct the energy densities of these holographic dark energy models, I assume the Hubble cutoff to be the IR limit.   
I find that the deceleration parameter undergoes a signature flipping at a redshift $z$ consistent with observations. The EoS parameter $\omega_{de}$ for both the HDE models exhibit quite contrasting dynamical behavior despite assuming values close to $-1$ at $z=0$ and therefore consistent with current observations. Next, I find the squared sound speed $c_{s}^{2}$ to be positive for the THDE model ensuring stability against perturbations, whereas for the RHDE model, $c_{s}^{2}<0$ implying instability against perturbations. Furthermore, I analyzed the evolutionary behavior of the EoS parameter of the HDE models by constructing the $\omega_{de}-\omega_{de}^{'}$ plane and find that the plane lies in the freezing region for the THDE model and in the thawing region for the RHDE model.   

\end{abstract}



\maketitle

\section{Introduction}
The fact that the universe at the present epoch is experiencing a period of accelerated expansion with an EoS parameter $\omega \simeq -1$ is well established and agreed upon \cite{note2}. However, there is no general consensus regarding the source which is fueling the acceleration. A plethora of well motivated theories have been reported which aim at sufficing this acceleration. Amongst them, Dark Energy seems to be the most plausible explanation. Many models of dark energy have been proposed in literature such as quintessence \cite{ratra}, K-essence \cite{picon}. Some alternate theories have also been reported such as the backreaction mechanism \cite{buchert}, voids \cite{pandey}, configurational entropy of the universe \cite{pandey2}, extra dimension \cite{milton}, entropic force \cite{easson} and entropy maximization \cite{pavon} and modified gravity \cite{note17}. \\
Holographic dark energy (HDE) is an intriguing candidate which can elegantly describe the current acceleration constructed from the holographic hypothesis \cite{note27} and is also compatible with multiple cosmological observations \cite{note30}. In these models, the horizon entropy is the most important aspect which upon altering, alters the HDE model drastically \cite{note}. \\
Gravitational systems can be studied through generalized statistical mechanics since gravity is a long range force \cite{note61,note64,note65,note66,note78}. Since, black hole entropy can be  obtained through the application of Tsallis statistics \cite{note61}, HDE models such as Tsallis HDE and R\'enyi HDE have recently been reported \cite{note1,note64,note65}. Readers are encouraged to see \cite{newHDE} for a comprehensive review of HDE models (also see \cite{reviewHDE} for some recent work in HDE). Amognst the two HDE models, without the interaction between the dark sectors, the RHDE constructed from the R\'enyi entropy is reported to be stable \cite{note65}. However, when an interaction between the dark sectors is presumed, both of these HDE models attains stability.\\
The fact that our universe may be filled with an interacting fluid exchanging energy between dark matter and dark energy cannot be ruled out completely \cite{note80to87,note88to90}. If such a fluid exists, it can lead to a possible solution to the coincidence problem \cite{note88to90,note91to96}. In this work I shall study the dynamics of interacting THDE and RHDE models in a FRW universe where the scale factor evolves with a composite power law-exponential (hybrid) form. To construct the energy density of these HDE models, I use the Hubble cutoff as the IR limit.\\
The paper is organized as follows: In Section \ref{sec2} I describe the hybrid expansion law and constrain the free parameters of the ansatz from the redshift of transition. In Section \ref{sec3} I outline the non-linear interaction between HDE models and cold dark matter. In Section \ref{sec4} I explain in detail the holographic dark energy models used in the work and study the behavior of dark energy EoS parameter for both the models. In Section \ref{sec5} I present the expressions of the sound speed for the HDE models to study their stability against perturbations. In Section \ref{sec6} I study the evolutionary behavior of the models by constructing the $\omega_{de}-\omega_{de}^{'}$ plane and in Section \ref{sec7} I present our results and conclusions.
\section{Hybrid Expansion Law}\label{sec2}
\begin{figure}[H]
\centering
\includegraphics[width=12 cm]{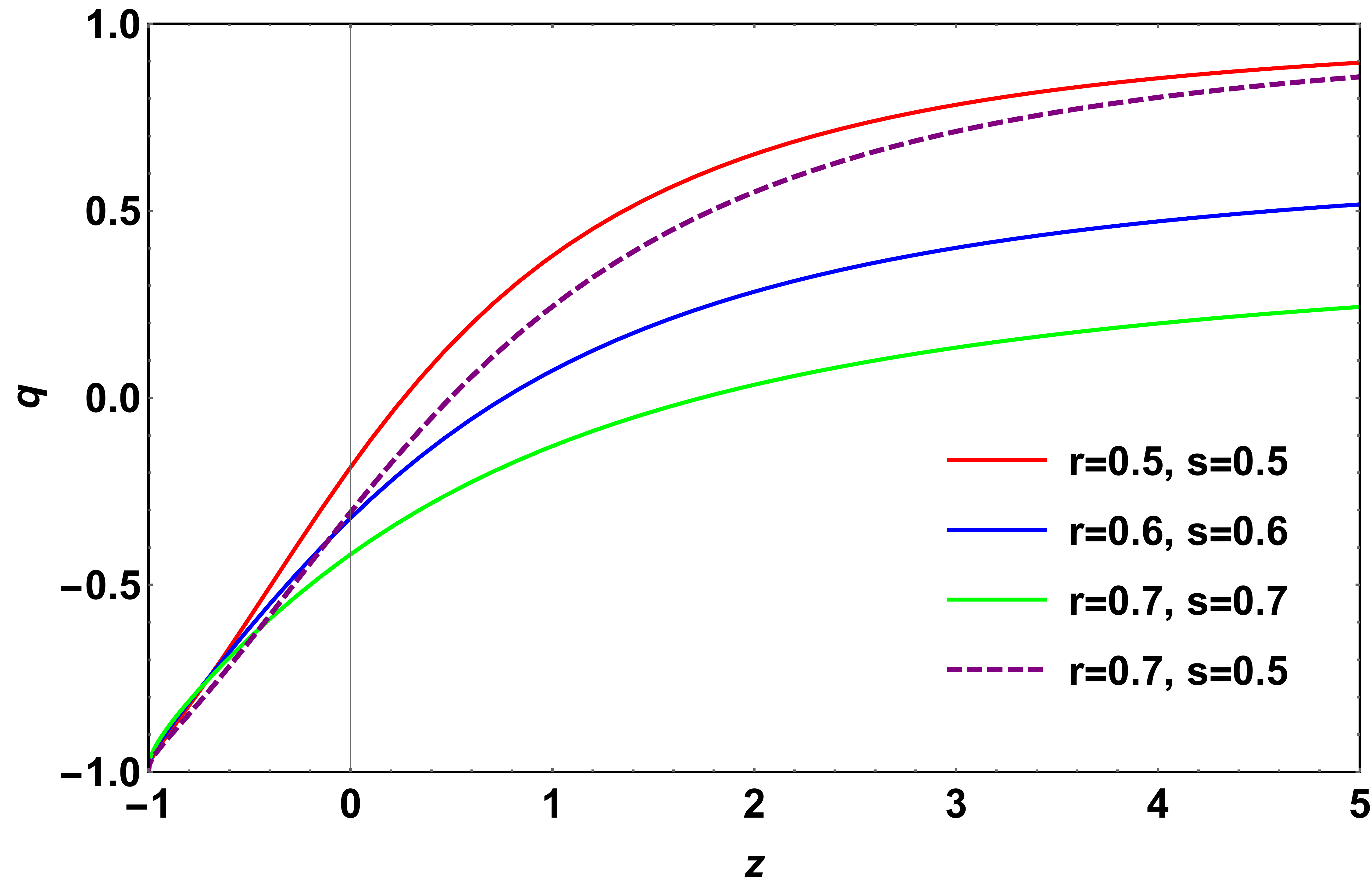}
\caption{Deceleration parameter as a function of redshift $z$.}
\label{f1}
\end{figure}

I shall now consider an ansatz of scale factor often refereed to as the hybrid expansion law (HEL). The motivation to employ such an ansatz is driven by the fact that it is both theoretically consistent and observationally viable \cite{non}. Additionally, in \cite{bhatta} the authors reported viable estimates of baryon to entropy ratio employing HEL in $f(R,T)$ gravity. The ansatz reduces to the usual power law and de Sitter solutions for special cases and also depicts the transition from a decelerated to an accelerated phase elegantly. \\
HEL has been tested against multiple cosmological observations such as BBN, CMB and BAO \cite{non35}. Remarkably, the cosmological parameters obtained from HEL lie well within $1-\sigma$ confidence level to that obtained from the standard $\Lambda$CDM model \cite{non35}. \\
Bearing that in mind, I consider the HEL to be of the form 
\begin{equation}\label{eq1}
a (t)= e ^{r t} t^{s}
\end{equation}  
where $r$ and $s$ are constants. The Hubble parameter and deceleration parameter reads 
\begin{equation}\label{eq2}
H (t)= r + \frac{s}{t}, \hspace{0.25in} q (t) = -1 + \frac{s}{(rt+s)^{2}}
\end{equation}
$r$ and $s$ needs to be carefully constrained so as to support the transition from a decelerated to an accelerated phase at a redshift of about $z_{r} \approx 0.55$.\\
From Eq. \ref{eq2}, it is obvious that the transition occurs at time $t=-\frac{s}{r} \pm \frac{\sqrt{s}}{r}$ with $0<s<1$. Since for negative $\frac{\sqrt{s}}{r}$, time becomes negative, I therefore infer that the transition must have taken place when $t=\frac{\sqrt{s} - s}{r}$.\\
using the relation $a(t) = \frac{1}{1+z}$, I arrive the following time-redshift relation 
\begin{equation}\label{eq3}
t= \frac{s \mathcal{W}\left[\frac{r (\frac{1}{1+z})^{\frac{1}{s}}}{s} \right] }{r}
\end{equation}
where $\mathcal{W}$ denotes the Lambert function. Employing Eq. \ref{eq3}, I plot the deceleration parameter as a function of redshift for different values of $r$ and $s$ in Fig. \ref{f1}. It is observed that when $r=s=0.5$, the transition occurs at a much smaller redshift which is unsuited with observations. Alternatively, when $r=s=0.7$, the signature flipping occurs at $z>1$ which again is ill-matched with observations. Fortuantely, I find that when $r=0.7$ and $s=0.5$ the transition occurs at $z_{r} \approx 0.50575$ in consistent with observations \cite{non36} and therefore throughout the analysis, I use this particular combination of $r$ and $s$.
\section{Non-Linear Interaction between dark matter and dark energy }\label{sec3}
If a mutual exchange of matter-energy take place between dark matter and dark energy, ther energy densities would assume a constant. Such scenarios could lead to a possible solution to the coincidence problem \cite{frame64}. For this work, I shall consider a non-gravitational, non-linear interaction between the dark sectors first studied in \cite{nonlinear}. The authors in \cite{nonlinear} argued that such type of interactions could lead to a future evolution which differs drastically from the de Sitter universe. \\
The general expression of the interaction term takes the form \cite{nonlinear}
\begin{equation}\label{eq4}
Q = 3 b H \rho_{eff}^{(i+j+k)}\rho_{dm}^{k}\rho_{de}^{-(j+k)}
\end{equation}
where $i,j,k$ and $b$ are constants, $\rho_{eff} =\rho_{dm}+\rho_{de} $ is the total energy density, $\rho_{dm}$ the energy density of cold dark matter and $\rho_{de}$ the same for dark energy.\\
Now the conservation equations for dark matter and dark energy assumes the form  
\begin{equation}\label{eq5}
\dot{\rho}_{dm}+3 H \rho_{dm} = -3 b H \rho_{eff}^{(i+j+k)}\rho_{dm}^{k}\rho_{de}^{-(j+k)}
\end{equation}
\begin{equation}\label{eq6}
\dot{\rho}_{de}+3 H (\rho_{de} +p_{de} ) = 3 b H \rho_{eff}^{(i+j+k)}\rho_{dm}^{k}\rho_{de}^{-(j+k)}
\end{equation}
where $p_{de}$ denote pressure of dark energy. Note that the strength of interaction depends on $i,j,$ and $k$. For example, for $(i,j,k) = (1,-1,1)$, I obtain $Q = 3 b H \rho_{dm}$ while for $(i,j,k) = (1,-1,1)$ I get $Q = 3 b H \rho_{de}$. For this work, I shall set $(i,j,k) = (1,-2,0)$ and therefore the interaction $Q$ assumes the form 
\begin{equation}\label{eq7}
Q = 3 b H \left(\frac{\rho_{de}^{2}}{\rho_{dm}+\rho_{de}} \right). 
\end{equation} 
Thence, Eqs. \ref{eq5} and \ref{eq6} become 
\begin{equation}\label{eq8}
\dot{\rho}_{dm}+3 H \rho_{dm} = -3 b H \left(\frac{\rho_{de}^{2}}{\rho_{dm}+\rho_{de}} \right)
\end{equation}
\begin{equation}\label{eq9}
\dot{\rho}_{de}+3 H (\rho_{de} +p_{de} ) =  3 b H \left(\frac{\rho_{de}^{2}}{\rho_{dm}+\rho_{de}} \right).
\end{equation}
The constant $b$ determines the direction in which energy flow take place. For positive $b$, energy flows from dark energy to dark matter and otherwise. Nonetheless, I shall assume positive $b$ in this work as it is favorable observationally and thermodynamically. 
\section{Holographic Dark Energy}\label{sec4}
The energy density of holographic dark energy can be written as    \cite{frame64} 
\begin{equation}
\rho_{de}=\frac{3 B^{2} m_{p}^{2}}{L^{2}},
\end{equation} 
where the constraint on the energy density of the holographic dark energy reads $L m_{p}^{2}\geq L^{3}\rho_{de}$. Here $m_{p}$ denote Planck mass, $B$ a dimensionless constant and $L$ the IR cutoff. Since the energy density depends on the choice of the IR cutoff, many such cutoffs have been proposed such as the Hubble cutoff, future event horizon cutoff, Ricci cutoff and the Granda-Oliveros cutoff. However, in this work I shall set the Hubble cutoff as the IR limit. \\
For a flat FRW background coupled with holographic dark energy and pressureless cold dark matter, the first Friedmann equation can be expressed as 
\begin{equation}\label{eq10}
H^{2}=\frac{\rho_{de}+\rho_{dm}}{3 m_{p}^{2}}.
\end{equation} 
I shall now study the dynamics two holographic dark energy models namely the Tsallis holographic dark energy and the Ren\'yi holographic dark energy in the following sections.
\subsection{Tsallis Holographic Dark Energy}
The horizon entropy of a black hole has been modified in \cite{note78} and assumes the form $S_{\delta}=\gamma A_{\delta}$, where $\delta$ denote the non-additivity parameter, $\gamma$ is a constant and $A=4 \pi L^{2}$ represent the surface area of the event horizon. Note that Bekenstein entropy can be obtained for $\delta=1$. In \cite{note27}, the authors reported a mutual relationship between the IR cutoff $L$, UV cutoff $\Gamma$ and the entropy $S$ as 
\begin{equation}
S^{3/4} \geq (L \Gamma)^{3}.
\end{equation} 
Upon substituting the value of $S$, yields 
\begin{equation}
\gamma 4 \pi ^{\delta} L^{(2\delta-4)} \geq \Gamma^{4}.
\end{equation} 
Now since the holographic dark energy $\rho_{de} \sim \Gamma^{4}$, the Tsallis holographic dark energy takes the final form \cite{note1}
\begin{equation}\label{eq11}
\rho_{de}=C L^{(2\delta-4)},
\end{equation}
where $C$ is a constant. Now to obtain the expression of the energy density of the THDE model prescribed by a non linear interaction and for a scale factor evolving with a composite power law-exponential (hybrid) form, I Substitute Eq. \ref{eq11} and Eq. \ref{eq2} in Eq. \ref{eq10} to obtain 
\begin{equation}\label{eq12}
\rho_{de} = C \left(r + \frac{s}{t}\right)^{4-2 \delta}. 
\end{equation}
The expression of pressure associated with this energy density can be obtained by substituting Eq. \ref{eq12} and Eq. \ref{eq2} in Eq. \ref{eq9} and reads 
\begin{equation}\label{eq13}
p_{de}=-C (r + \frac{s}{t})^{(4 - 2 \delta)} + 
 \frac{1}{3} (r + \frac{s}{t}) \left[ C^2 b( r + \frac{s}{t})^{(7 - 4 \delta)} + \left( 
   \frac{C s (r + \frac{s}{t})^{(3 - 2 \delta)} (4 - 2 \delta))}{t^{2}} \right)\right].  
\end{equation} 
The EoS parameter $\omega_{de} =p_{de}/ \rho_{de} $ reads 
\begin{equation}\label{eq14}
\omega_{de} = \frac{C b \left( r + \frac{s}{t}\right)^{-2 \delta} (s + r t)^{4} + t^{2}(-3 t^{2} - 2 s (\delta - 2))}{3 t^{4}}.
\end{equation}
\begin{figure}[H]
\centering
\includegraphics[width=12 cm]{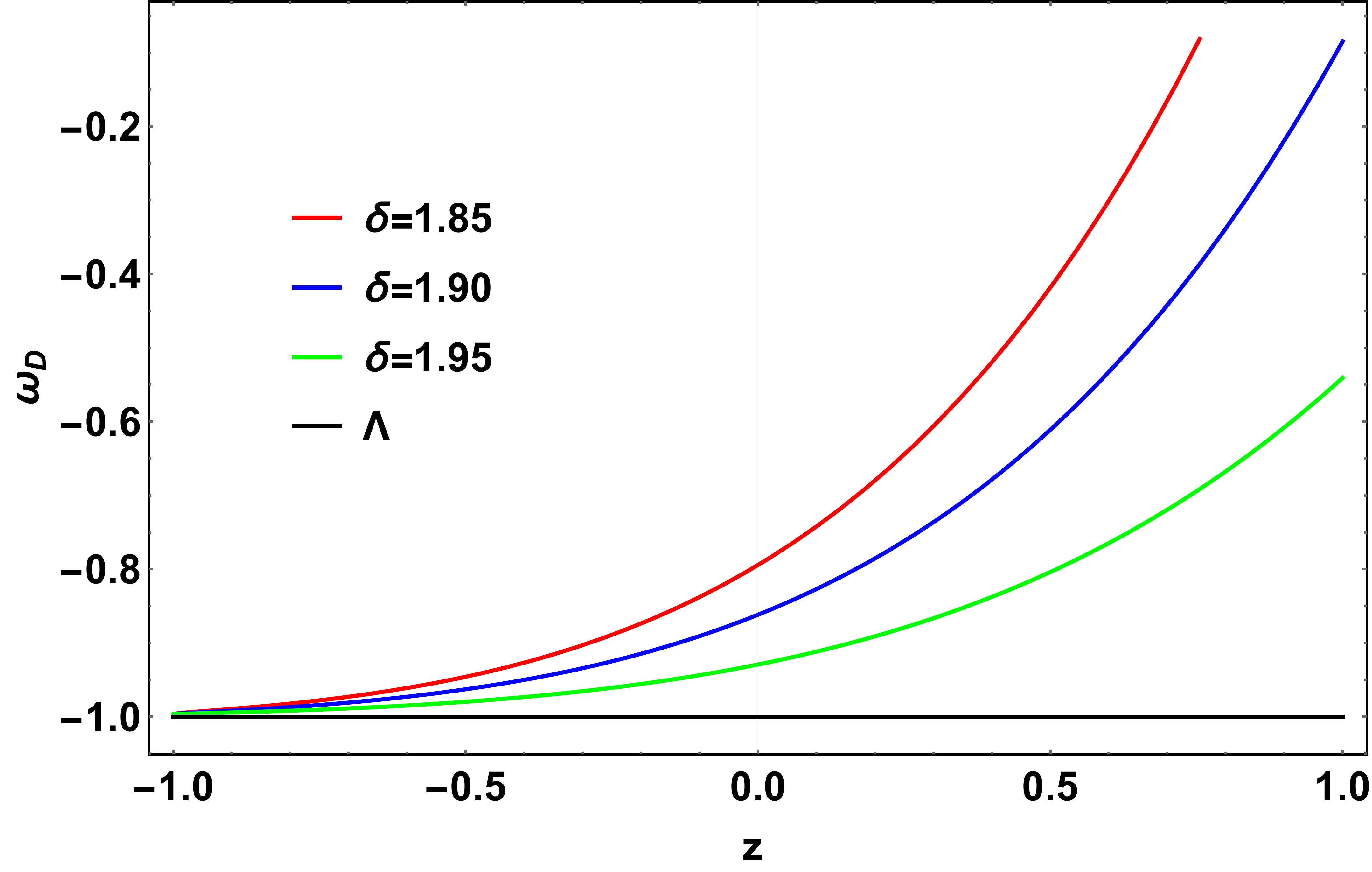}
\caption{EoS parameter as a function of redshift $z$ with $r=0.7$, $s=0.5$, $C=1$ and $b=0.01$.}
\label{f2}
\end{figure}
In Fig. \ref{f2} I show the evolution of THDE dark energy EoS parameter $\omega_{de}$ for some rational values of Tsallis parameter $\delta$. For all $\delta$, $\omega_{de}$ remains in the Quintessence region and approaches the Phantom divide line in future (i.e, for $z<0$). It can also be noted that as $\delta$ decreases the profiles shift towards higher values at redshift $z=0$ and beyond. The solid black line corresponds to the dark energy EoS parameter of the cosmological constant $\Lambda$ which assumes a value of $-1$ at all times. 
\subsection{R\'enyi Holographic Dark Energy}
R\'enyi entropy can be written as \cite{note65}
\begin{equation}
S=\frac{1}{\delta} \log \left[1+\pi \delta L^{2} \right]. 
\end{equation}
Now considering $\rho_{de} d V \propto T dS$, where $T$ and $V$ denote the volume and temperature of the system, the expression of  R\'enyi holographic dark energy assumes the form \cite{note65}
\begin{equation}\label{eq15}
\rho_{de} = \frac{3 D^{2}}{8 \pi L^{2}}\left(1+ \pi \delta L^{2} \right) ^{-1}
\end{equation}
where $D$ is a constant. Similar to the previous section, I shall substitute Eq. \ref{eq15} and Eq. \ref{eq2} in Eq. \ref{eq9}, to obtain the expression of the energy density for the RHDE model prescribed by a non linear interaction and for a scale factor evolving with a composite power law-exponential (hybrid) form and reads
\begin{equation}\label{eq16}
\rho_{de}= \frac{3 D^{2} (s + r t)^{4}}{8 \pi t^{2} \left((s + r t)^{2} + \pi t^{2}\delta \right) }.
\end{equation}
The expression of pressure and EoS parameter for the RHDE model can now be obtained in a similar way and reads respectively as
\begin{equation}\label{eq17}
p_{de}=\frac{D^{2}(s + r t)^{4} \left(3 b D^{2}(s + r t)^{4} + 8 \pi (s + r t)^{2} (2s-3t^{2})  + 8 \pi^{2}t^{2}(4s-3t^{2})\delta\right) }{64 \pi^{2}t^{4}\left((s + r t)^{2}+\pi t^{2} \delta\right)^{2} }
\end{equation}
\begin{equation}\label{eq18}
\omega_{de} = \frac{\left(3 b D^{2}(s + r t)^{4} + 8 \pi (s + r t)^{2} (2s-3 t^{2})  + 8 \pi^{2}t^{2}(4s-3 t^{2}) \delta \right)} {24 \pi t^{2}\left((s + r t)^{2}+\pi t^{2} \delta\right)}.
\end{equation}
\begin{figure}[H]
\centering
\includegraphics[width=12 cm]{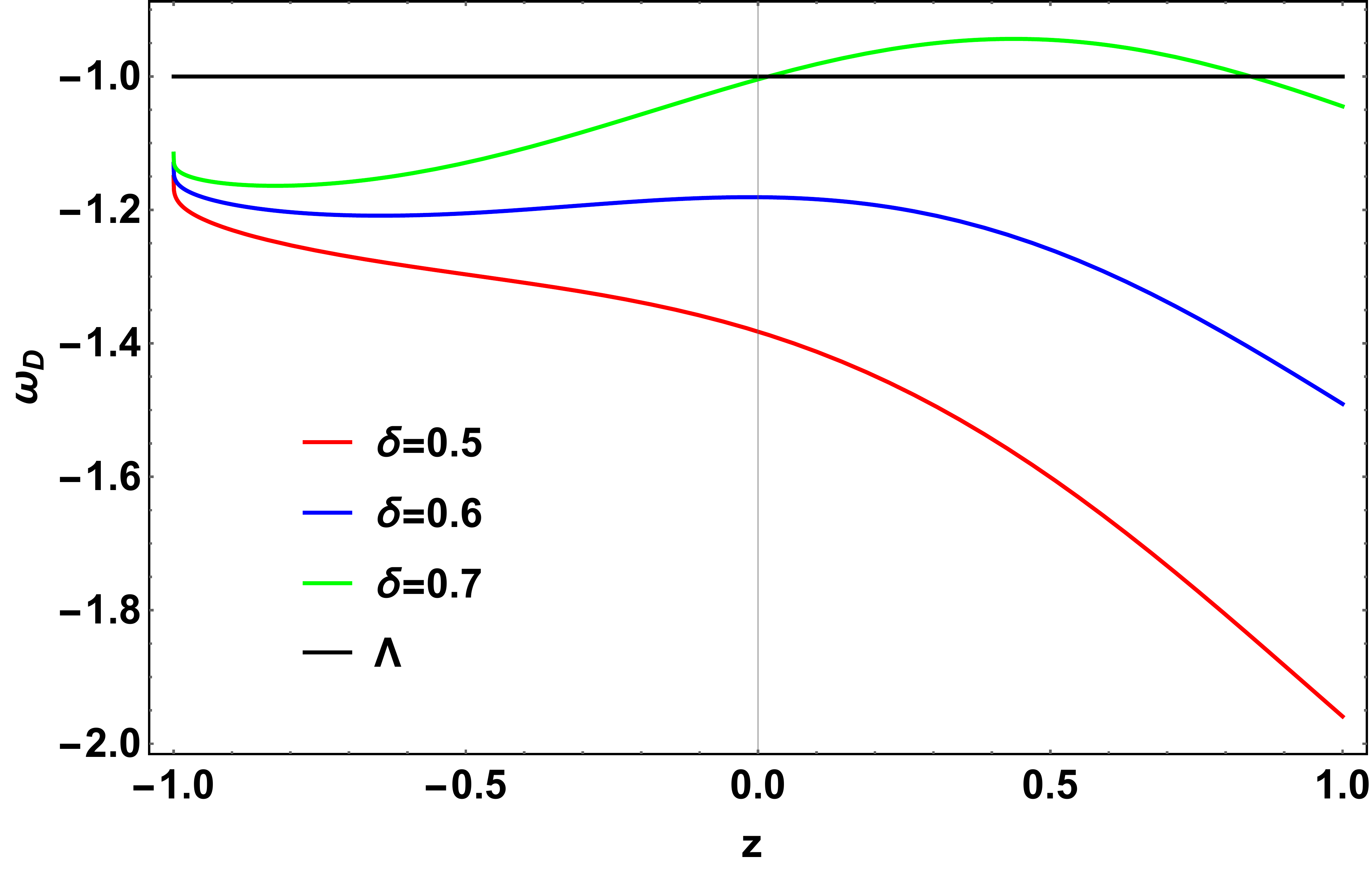}
\caption{EoS parameter as a function of redshift $z$ with $r=0.7$, $s=0.5$, $D=-1$ and $b=3.65$.}
\label{f3}
\end{figure}
In Fig. \ref{f3} I show the evolution of RHDE dark energy EoS parameter $\omega_{de}$ for some values of R\'enyi parameter $\delta$. From the figure it is clear that for $\delta =0.5$ and $0.6$, $\omega_{de}$ lies in the Phantom region for the illustrated redshift range. Interestingly, for $\delta=0.7$, $\omega_{de}$ assume the value of $-1$ at $z=0$ in harmony with observations. The profile then enters the Quintessence region for some time and then reverts back to being Phantom in nature. The solid black line corresponds to the dark energy EoS parameter of the cosmological constant $\Lambda$ which assumes a value of $-1$ at all times. 
\section{Evolution of Sound Speed}\label{sec5}
The square of sound speed ($c_{s}^{2}$) is a useful parameter to understand the stability of a dark energy model against perturbations. If $c_{s}^{2}>0$, the model is stable and otherwise unstable \cite{peebles}. The expression of $c_{s}^{2}$ reads 
\begin{equation}\label{eq19}
c_{s}^{2} = \frac{\partial p}{\partial \rho}.
\end{equation}
\subsection{For Tsallis Holographic Dark Energy}
For the THDE model, the expression of $c_{s}^{2}$ reads 
\begin{equation}\label{eq20}
c_{s}^{2} =\frac{2 C b (r + \frac{s}{t})^{-2 \delta} (s + r t)^{4} + t^{2}\left( (2r-3t)t - 2 s (\delta - 3)\right) }{3 t^{4}}.
\end{equation}
\begin{figure}[H]
\centering
\includegraphics[width=12 cm]{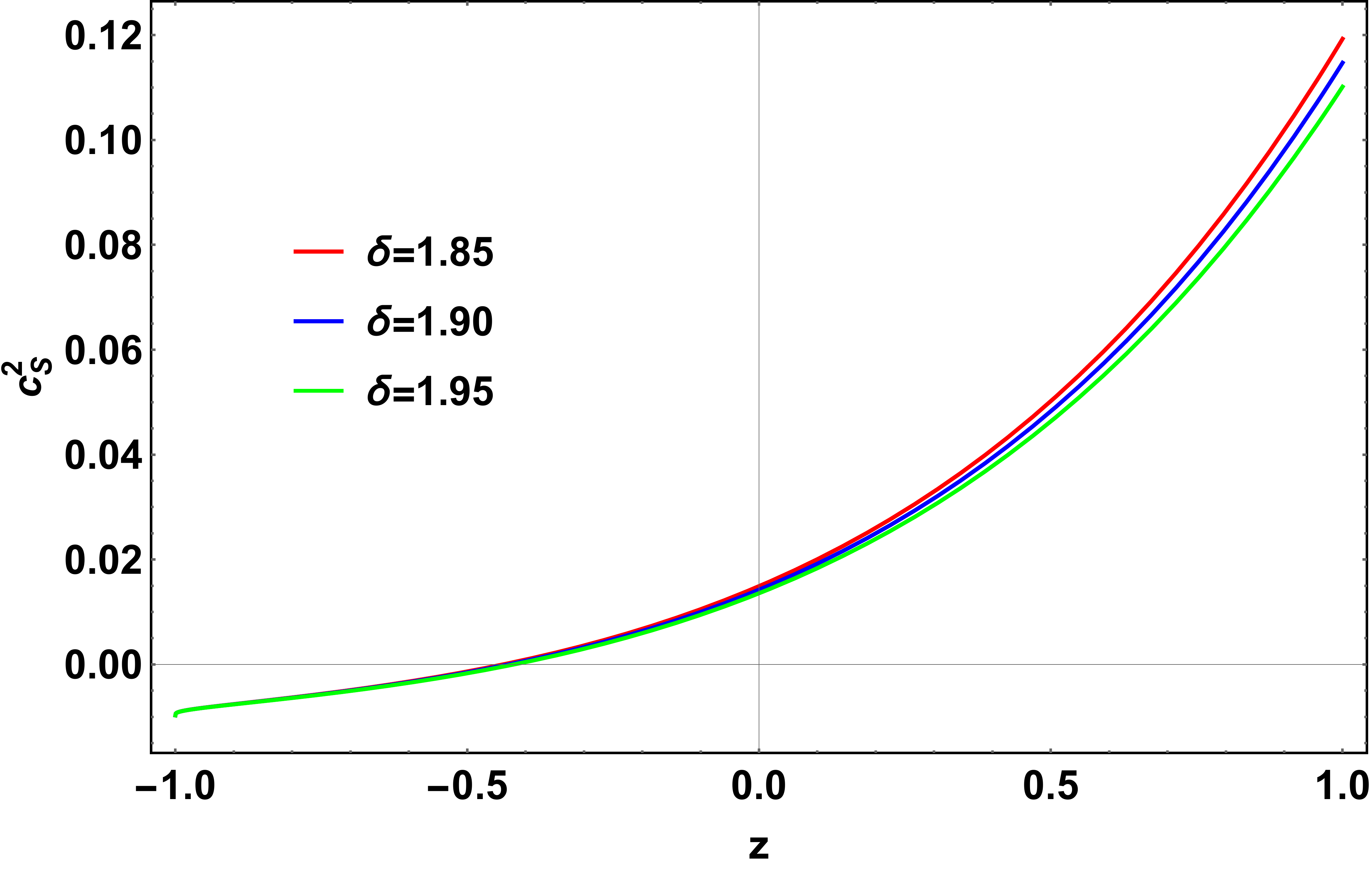}
\caption{Squared sound speed as a function of redshift $z$ with $r=0.7$, $s=0.5$, $C=1$ and $b=0.01$.}
\label{f4}
\end{figure}
In Fig. \ref{f4} I show the evolution of squared sound speed ($c_{s}^{2}$) for the THDE model for some values of Tsallis parameter $\delta$. It is evident that for all $\delta$, $c_{s}^{2} >0$ and therefore ensures stability against perturbations. However in future (particularly for $z\lesssim -0.5$), the model becomes unstable. It can be noted that $c_{s}^{2}$ is not very sensitive to $\delta$. 
\subsection{For R\'enyi Holographic Dark Energy}
For the RHDE model, the expression of $c_{s}^{2}$ reads 
\begin{equation}\label{eq21}
c_{s}^{2} =\frac{3 b D^{2}(s+rt)^{6} + 12 \pi^{2}t^{2}(s+rt)^{2}\left[ 4s+(2r-3t)t\right] \delta + 8 \pi^{3}t^{4}\left[ 6s+(2r-3t)t\right] \delta^{2}+2\pi (s+rt)^{4}(8s+t)(4r-6t+3bD^{2}t\delta)}{12 \pi t^{2}\left[(s+rt)^{4}+3 \pi t^{2} (s+rt)^{2}\delta +2 \pi^{2}t^{4}\delta^{2} \right] }.
\end{equation}
\begin{figure}[H]
\centering
\includegraphics[width=12 cm]{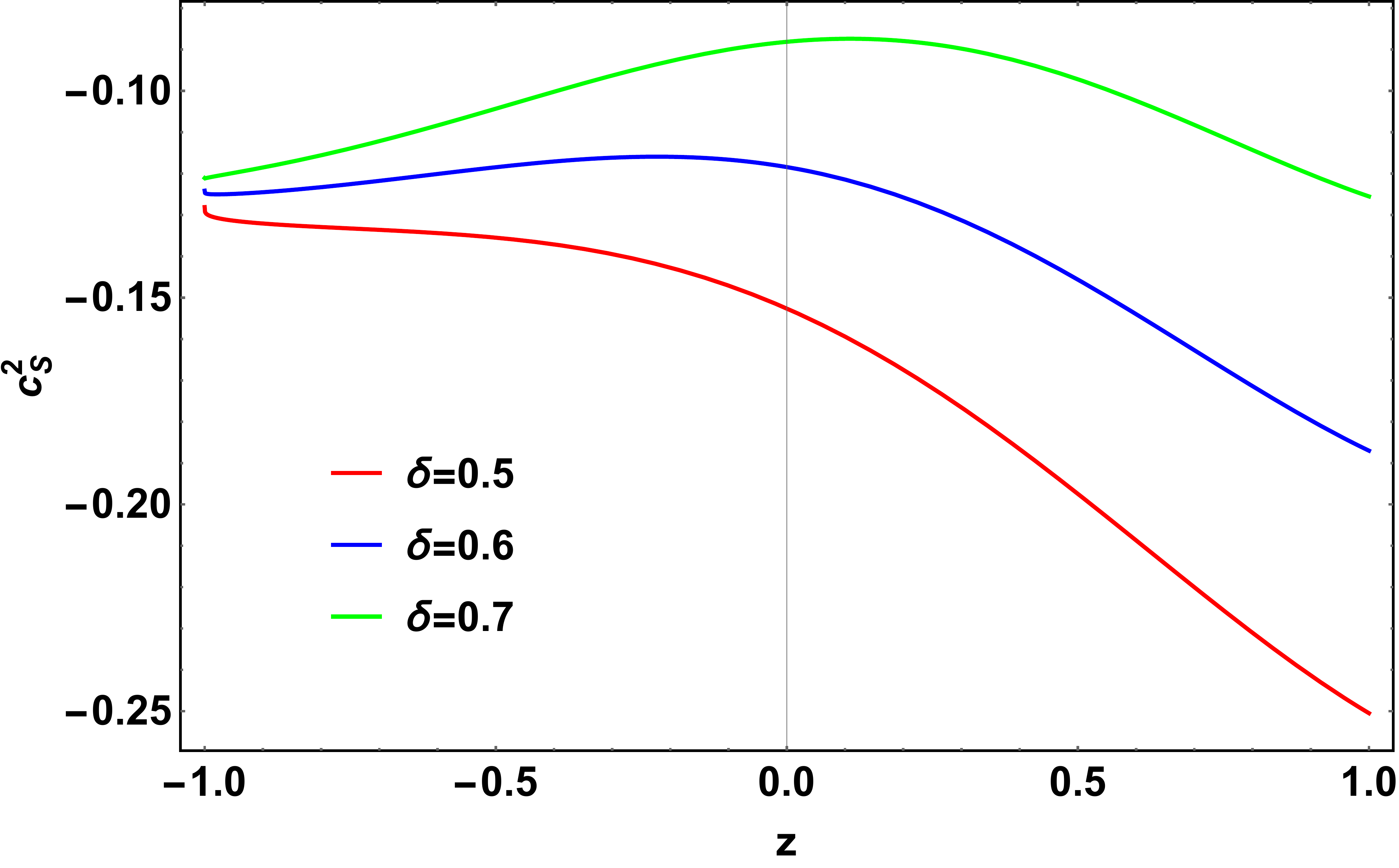}
\caption{Squared sound speed as a function of $z$ with $r=0.7$, $s=0.5$, $D=-1$, and $b=3.65$.}
\label{f5}
\end{figure}
In Fig. \ref{f5} I show the evolution of squared sound speed ($c_{s}^{2}$) for the RHDE model for some values of R\'enyi parameter $\delta$. Unlike the THDE model, the RHDE model is unstable against perturbations since $c_{s}^{2}<0$ for all $\delta$. Moreover, $c_{s}^{2}$ is highly sensitive to $\delta$ unlike the THDE model. It can also be noted that negativity of $c_{s}^{2}$ increases as $\delta$ is decreased.   
\section{$\omega_{de}-\omega_{de}^{'}$ Plane}\label{sec6}
To explain the dynamical properties of dark energy models, \cite{frame70} proposed the $\omega-\omega^{'}$ plane where $\omega_{de}^{'}$ represent the derivative of $\omega_{de}$ with respect to lna. The plane is sectioned into two parts: The thawing region where ($\omega_{de} <0, \omega_{de}^{'} >0$) and the freezing region where ($\omega_{de} <0, \omega_{de}^{'}<0$). \\
\subsection{For Tsallis Holographic Dark Energy}
For the THDE model, the expression of $\omega_{de}^{'}$ reads
\begin{equation}\label{eq22}
\omega_{de}^{'} = \frac{d \omega_{de} }{\text{dlna}} = \frac{2 r^{2} (\delta-2)\left[2 (\frac{t r}{s})^{2} + C d r^{2}s (r + \frac{s}{t})^{-2 \delta} ( 1 + \frac{tr}{s})^{3} \right] }{3 s^{2}(\frac{t r}{s})^{4}(1+\frac{t r}{s})}.
\end{equation} 
\begin{figure}[H]
\centering
\includegraphics[width=12 cm,height=8 cm]{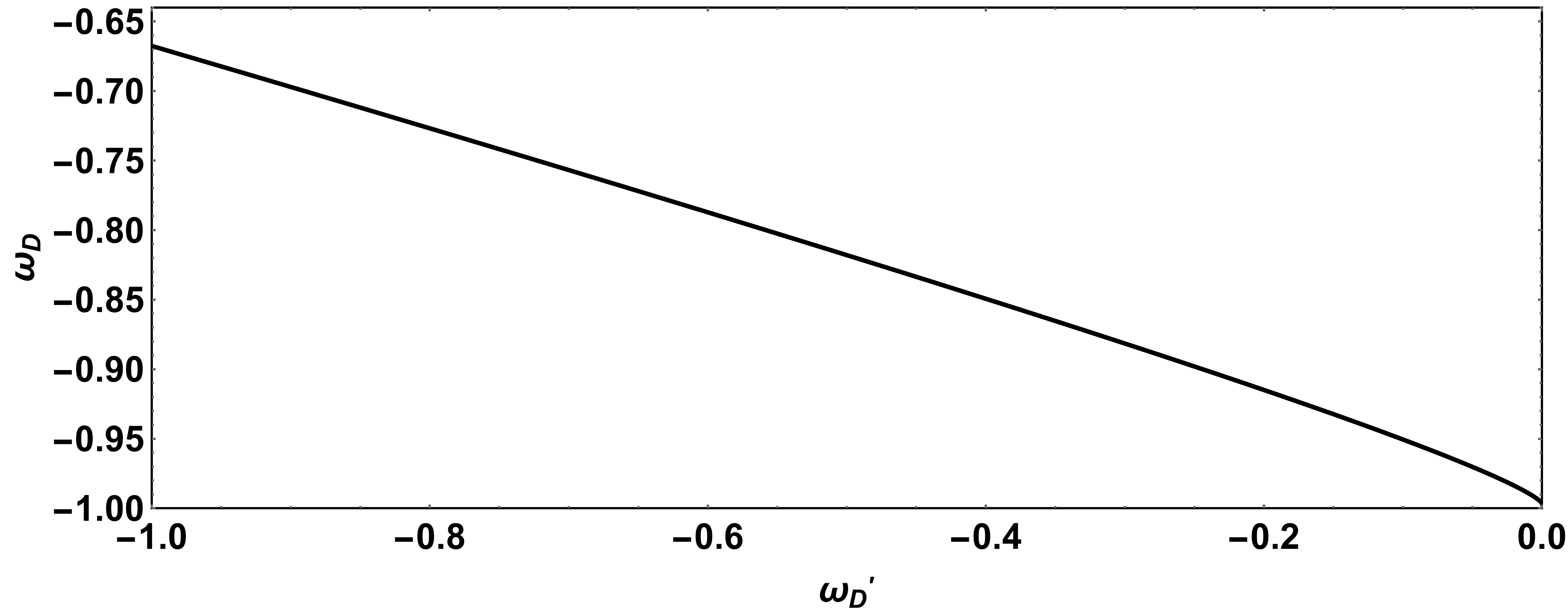}
\caption{$\omega_{de}-\omega_{de}^{'}$ plane with $r=0.7$, $s=0.5$, $C=1,b=0.01,$ and $\delta=1.95$.}
\label{f6}
\end{figure}
\subsection{For R\'enyi Holographic Dark Energy}
For the RHDE model, the expression of $\omega_{de}^{'}$ reads 
\begin{equation}\label{eq23}
\omega_{de}^{'}  = \frac{d \omega_{de} }{\text{dlna}} =   \frac{\left[\Sigma \left( \Pi \left(\Theta (16\pi + 3 b D^{2}s)\delta-2r^{2}(r^{2}(32\pi + 15 b D^{2}s))+\pi (40\pi + 9 b D^{2}s)\frac{tr}{s}-\Theta(8\pi + 3 b D^{2}s)\delta + \Upsilon(\frac{tr}{s})^{3} \right)\right)   \right] }{\left[12 \pi s^{2} (\frac{tr}{s})^{2}(1+\frac{tr}{s}) (r^{2}+2r^{2}(\frac{tr}{s}) + (r^{2}+\pi \delta) \frac{tr}{s})^{2}\right]  }
\end{equation}
where 
\begin{equation}
\Sigma= -r^{6}(16\pi +3 b D^{2}s) +r^{2}\frac{tr}{s}
\end{equation}
\begin{equation}
\Pi= -r^{4}(64 \pi + 15 b D^{2}s)+ \frac{tr}{s}
\end{equation}
\begin{equation}
\Theta=-6r^{4}(16\pi + 5 b D^{2}s) - 2 \pi r^{2}
\end{equation}
\begin{equation}
\Upsilon=32 \pi^{3}\delta^{2}(\frac{tr}{s})^{2} -3bD^{2}r^{2}s(r^{2}+2 \pi \delta).
\end{equation}
\begin{figure}[H]
\centering
\includegraphics[width=12 cm,height=7.5 cm]{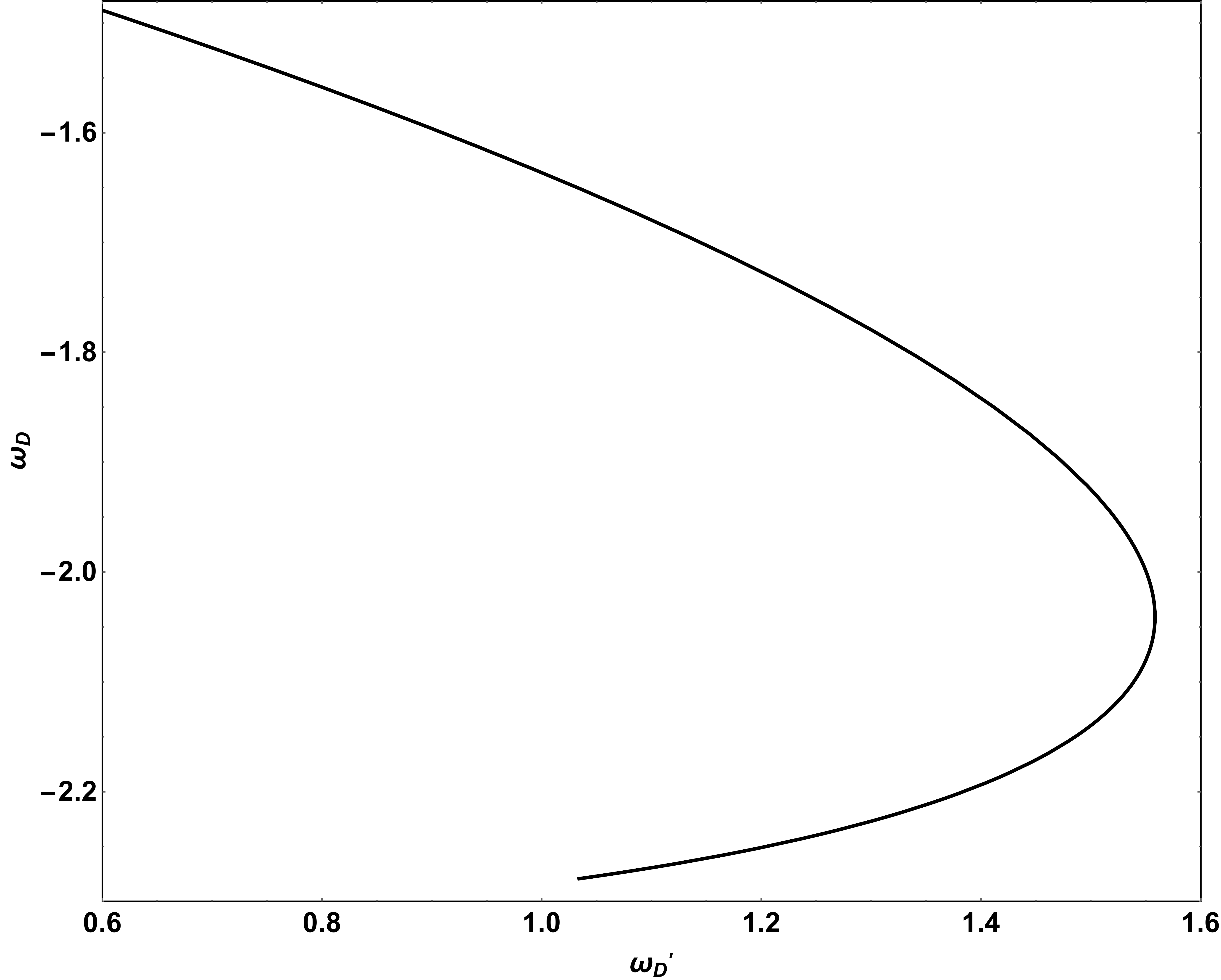}
\caption{$\omega_{de}-\omega_{de}^{'}$ plane with $r=0.7$, $s=0.5$, $D=-1,b=3.65,\delta=0.6$.}
\label{f7}
\end{figure}
\section{Results \& Conclusions}\label{sec7}
Holographic dark energy (HDE) models are a class of dark energy models constructed from the holographic hypothesis and is becoming widely popular owing to its elegant explanation of the current acceleration of the universe in harmony with multiple cosmological observations. In HDE models, the horizon entropy plays the most crucial part in constructing the dark energy density. Different IR cutoffs result in different dark energy densities which are then confronted with observations. \\
In this work I investigated the dynamics of two recently proposed holographic dark energy models namely the Tsallis holographic dark energy (THDE) and the R\'enyi holographic dark energy (RHDE) in the FRW spacetime with the scale factor evolving as a composite power law-exponential (hybrid) form. I also assumed a non-linear interaction between dark matter and dark energy. To construct the expressions of the energy densities of the HDE models, I assumed the Hubble cutoff to be the IR limit. The results can be summarized as follows:
\begin{itemize}
\item The deceleration parameter undergoes a signature flipping for suitable combinations of $r$ and $s$. For $r=0.7$ and $s=0.5$ the transition occurs at $z_{r} \approx 0.50575$ in consistent with observations \cite{non36}. 
\item For all values of Tsallis parameter $\delta$, $\omega_{de}$ of the THDE model remains in the Quintessence region and approaches the Phantom divide line in future (i.e, for $z<0$). I also find that as $\delta$ decreases the profiles shift towards higher values at redshift $z=0$ and beyond.\\
In the case of RHDE model, I find that when the R\'enyi parameter $\delta =0.5$ and $0.6$, $\omega_{de}$ lies in the Phantom region. Nontheless, when $\delta=0.7$, $\omega_{de}$ assume the value of $-1$ at $z=0$ in harmony with observations. The profile then enters the Quintessence region for some time and then reverts back to being Phantom in nature.
\item Next, for the THDE model, the squared sound speed $c_{s}^{2} >0$ for all $\delta$ and therefore ensures stability against perturbations. Interestingly, for $z\lesssim -0.5$, the model becomes unstable. I also find that $c_{s}^{2}$ is not very sensitive to $\delta$. \\
The squared sound speed for the RHDE model is negative implying instability against perturbations. Moreover, $c_{s}^{2}$ is highly sensitive to $\delta$. It can also be noted that as $\delta$ decreases the negativity of $c_{s}^{2}$ increases.   
\item Finally I analyze the evolutionary behavior of the EoS parameter of the HDE models by constructing the $\omega_{de}-\omega_{de}^{'}$ plane and find that $\omega_{de}^{'}<0$ for $\omega_{de}<0$ for the THDE model implying the plane to lie in the freezing region while for the RHDE model, $\omega_{de}^{'}>0$ for $\omega_{de}<0$ entailing the plane to lie in the thawing region.   
\end{itemize} 

\section*{Acknowledgments}
I thank DST, New-Delhi, Government of India for the provisional INSPIRE fellowship selection [Code: DST/INSPIRE/03/2019/003141]. I also thank the anonymous referee for his/her constructive comments and useful criticisms which greatly improved the quality and presentation of the work.

\end{document}